\documentstyle[11pt,newpasp,epsf,twoside]{article}
\def\edcomment#1{\iffalse\marginpar{\raggedright\sl#1\/}\else\relax\fi}
\reversemarginpar

\begin{document}
\title{Magnetic CVs in the UCT CCD CV Survey}

\author{Patrick A. Woudt and Brian Warner} 

\affil{Department of Astronomy, University of Cape Town, Rondebosch 7700, South Africa}

\begin{abstract}
An overview is given of all the magnetic CVs found in the UCT CCD CV Survey (Woudt \& Warner
2001, 2002, 2003a). We have identified eight new candidate Intermediate Polars (IP), of which six are
classical novae (RR Cha, DD Cir, AP Cru, V697 Sco, V373 Sct, and RX\,J1039.7-0507). The two other 
candidate IPs are Aqr1 (2236+0052) and RX\,J0944.5+0357. In addition, there are two probable Polars,
namely V351 Pup (= Nova Puppis 1991) and FIRST J102347.6+003841.
\end{abstract}

\section{The UCT CCD CV Survey}

The UCT CCD CV Survey is an ongoing high-speed photometric survey of faint Cataclysmic Variable (CV) stars
(Woudt \& Warner 2001, 2002, 2003a) down to a magnitude limit of V $\sim$ 21. For this survey, we use the
1.9-m and 1.0-m telescopes of the South African Astronomical Observatory at the Sutherland site, in 
combination with the University of Cape Town (UCT) CCD camera (O'Donoghue 1995). In the past three years,
an intensive observational campaign -- mainly focused on old novae -- has resulted in new orbital periods
for 18 CVs, 12 of which are old southern novae.

\section{Candidate Intermediate Polars}

The objects listed in this section are candidate Intermediate Polars. In these systems we have either seen
a second (consistent) photometric period in addition to the orbital modulation, e.g., DD Cir, or we have identified multiple 
photometric periodicities associated with a spin period (or its harmonic) 
and its orbital sidebands, e.g., RX\,J1039.7-0507. 
In the former case, confirmation of the suspected spin period in X-ray, or optical sidebands, are needed to confirm its 
classification, whereas in the latter scenario, the orbital sidebands already secure the IP nature of the CV.

We will list each of the eight candidate IPs with a short description of their main characteristics,
a summary is given in Table 1.\\

{\bf Aqr1:} Listed in Downes et al.~(2001) as Aqr1, this object is also known as 2236+0052 and was 
identified spectroscopically as a CV by Berg et al.~(1992) in a bright QSO survey. We have observed
Aqr1 during four nights in 2002 November; the light curves of Aqr1 (for two of the nights) 
are shown in Fig.~1.
The Fourier Transform (FT) of the combined data set, after subtracting the mean of each run, is shown in 
Fig.~2. There is an unambiguous peak at $\omega$ = 2477.1 $\mu$Hz, which we infer as the spin frequency.
In addition, there are two prominent sidebands ($\omega - \Omega$ = 2391.2 $\mu$Hz and 
$\omega + \Omega$ = 2563.2 $\mu$Hz) 
-- see also the blow up in the upper panel of Fig.~2 -- approximately equally spaced on either side 
of $\omega$; the separation is 85.9 $\mu$Hz and 86.1 $\mu$Hz, respectively, for $\omega - \Omega$ and $\omega + \Omega$.
The orbital frequency ($\Omega$) cannot be determined unambiguously from these data, as the
one day alias (towards lower frequency, i.e. a larger orbital period) cannot be excluded. 
The peak at low frequency in the lower panel of Fig.~2 does not correspond to the orbital frequency,
nor one of its aliases. A more thorough
study of Aqr1 is needed in order to determine its orbital period. The classification as an
Intermediate Polar, however, is secure due to the presence of the orbital sidebands.\\

\begin{figure}[t]
\plotfiddle{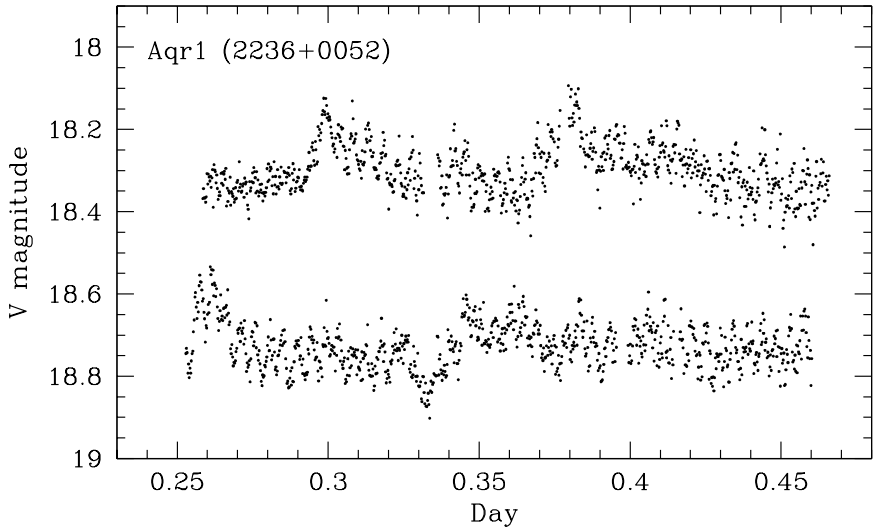}{7.2cm}{0}{65}{65}{-195}{-230}
\caption{The light curves of Aqr1 (2236+0052) obtained on 2002 November 1 (upper light curve) and 
November 2 (lower light curve, displaced downwards by 0.4 mag).}
\label{woudtfig1}
\end{figure}

\begin{figure}[t]
\plotfiddle{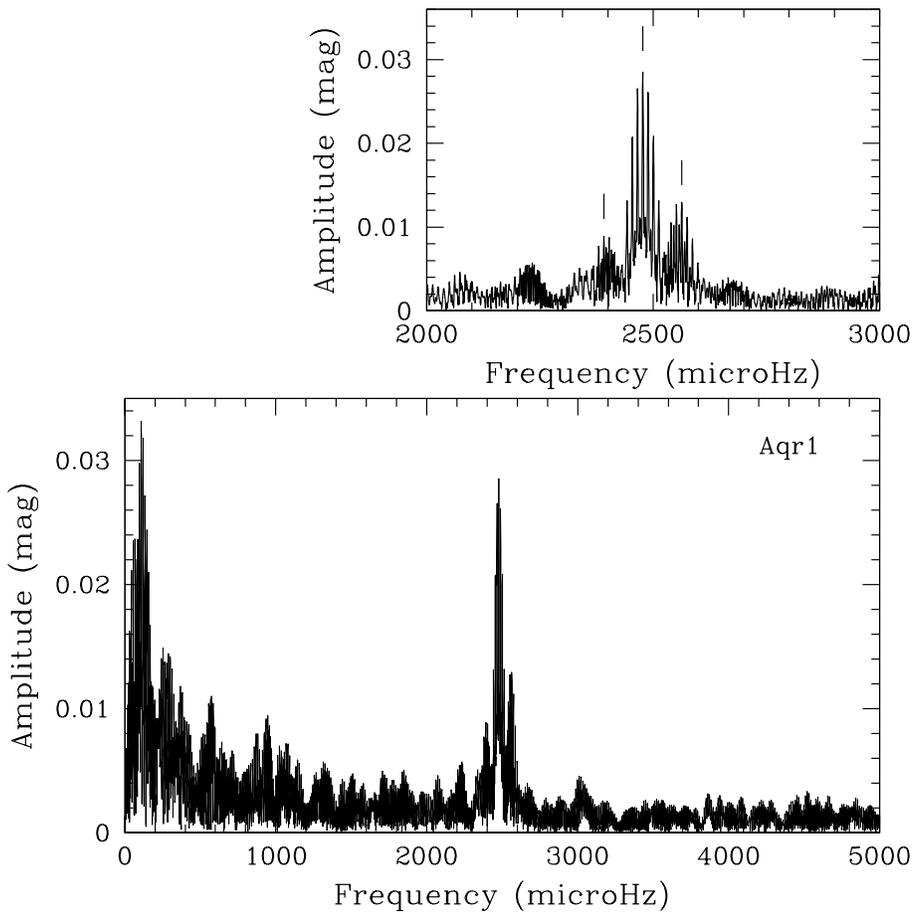}{12.0cm}{0}{65}{65}{-195}{-105}
\caption{The Fourier transform of Aqr1 (2236+0052) of the 2002 November data.}
\label{woudtfig2}
\end{figure}

{\bf RR Cha:} RR Cha (Nova Chamaeleontis 1953) is an eclipsing system (P$_{orb}$ = 3.362 h) which
shows both positive and negative superhumps (Woudt \& Warner 2002). In two independent data sets (2001 February
and 2001 May) we identified a photometric periodicity at 1950 s, suggesting that this might be either the spin
period of the white dwarf, or the reprocessed period. It is not detected in the ROSAT X-ray survey (G.~Israel, private
communication), but recent polarimetric observations of RR Cha (Rodr\'\i{guez}-Gil \& Potter, in prep.) 
revealed circular polarisation at the 1950 s periodicity, supporting the classification as an IP.\\

{\bf DD Cir:} DD Cir (Nova Circini 1999) is an eclipsing system (Woudt \& Warner 2003a) with P$_{orb}$ = 2.339 h
and eclipses $\sim$ 0.6 mag deep. It has a mean magnitude out of eclipse of V $\sim$ 20.1 mag. The light curve of
DD Cir reveals a number of interesting features: (i) there is a reflection effect of amplitude $\sim$ 0.3 mag, indicative of a recent
nova outburst in a system where the disc does not dominate the 
luminosity of the system (a short orbital period and high inclination both favour such reflection effect), 
(ii) there is evidence for a secondary eclipse, and (iii) there is a consistent photometric modulation at $\sim$ 670 s which 
is either the spin period of the white dwarf, or its orbital sideband.\\

{\bf AP Cru:} AP Cru was Nova Crucis 1936 and is currently at V $\sim$ 18. An uncharacteristic CV spectrum (absorption
spectrum of a K7 star with strong narrow H$\alpha$ emission superimposed) was observed for AP Cru (Munari \& Zwitter 1998), 
when the system was 0.6 -- 0.9 mag fainter than our observations. We suggest that the star was at a very
low (or zero) mass transfer rate then. We find an orbital period of 5.12 h, and in the FT there is a persistent
signal at 1837 s, suggestive of AP Cru being an IP (Woudt \& Warner 2002). 
There is no X-ray emission observed from AP Cru; its large distance
and low Galactic latitude (2$^\circ$) make it improbable of being detected in X-ray.  \\

{\bf V697 Sco:} We observed V697 Sco (Nova Scorpii 1941) during two nights in 2001 May, and four nights in 2002
June (Warner \& Woudt 2002). It is faint (V $\sim$ 20) and located in a star-crowded region. The orbital frequency
and its first harmonic are very distinctly present in the FT (P$_{orb}$ = 4.49 h), but after prewhitening
at these frequencies, a set of peaks remain which can be explained in terms of $\omega$ (and its harmonics),
and orbital sidebands. The spin period (2$\pi$/$\omega$) of V697 Sco is 3.31 h, and hence the model that emerges
for V697 Sco is that of an asynchronous magnetic rotator, similar to IPs like EX Hya and V1025 Cen, but at a much
larger P$_{orb}$. For a detailed discussion of V697 Sco, see Warner \& Woudt (2002).\\

{\bf V373 Sct:} V373 Sct (Nova Scuti 1975) was observed by us on three nights in 2002 April. The light curves show
great activity but no orbital modulation. The only significant modulation in any of the three light
curves was a narrow spike at 258.3 s seen in one of the runs (S6361, see Woudt \& Warner 2003a). More observations
are needed in the optical and X-ray regions to confirm its possible IP classification. \\

{\bf RX\,J0944.5+0357:} Spectroscopic observations of RX\,J0944.5+0357 (hereafter RXJ0944) 
by Jiang et al.~(2000) showed H\,I and He\,I emission lines typical of a CV. Additional spectroscopy
by Mennickent et al.~(2002) confirmed the previously determined (spectroscopic) orbital period
of 3.581 h reported to them by Thorstensen \& Fenton. High speed photometry of RXJ0944 by Mennickent et
al.~(2002) revealed large amplitude variations ($\sim$ 0.5 mag), but no coherent signal was reported in
their data. 

Our high speed photometry of RXJ0944, taken in 2002 March and April, revealed a repetitive brightness modulation
with a period of $\sim$ 2000 s. This feature has a double humped profile, with the two humps varying independently
and rapidly in amplitude. One of our light curves illustrating this most clearly is shown in Fig.~3. The best
period (from the Fourier Transform of the entire data set) is 2162 s and is shown in Fig.~3 as the lower vertical
bars. RXJ0944 is quantitatively similar to canonical IPs such as FO Aqr and TV Col, however, it most resembles
YY Dra (e.g., the double-humped light curve). A more extensive discussion of our observations can be found in
Woudt \& Warner (2003b).\\

\begin{figure}[t]
\plotfiddle{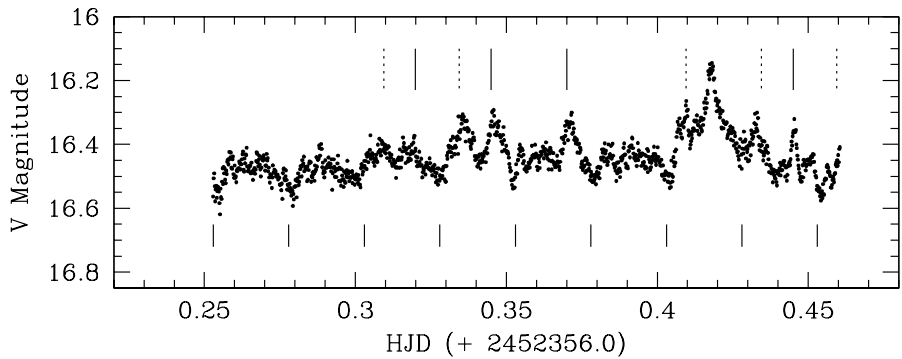}{5.0cm}{0}{65}{65}{-195}{-225}
\caption{The light curve of RX\,J0944.5+0357 obtained on 2002 March 22. The 2162 s periodicity
is marked by the lower vertical bars; the pairs of peaks of variable amplitude are marked by
the upper bars (dotted bar for the first peak, solid bar for the second peak).}
\label{woudtfig3}
\end{figure}

{\bf RX\,J1039.7-0507:} RX\,J1039.7-0507 (hereafter RXJ1039) was identified by Appenzeller et al. (1998)
as a CV at V = 18.5. Our photometry of RXJ1039 (Woudt \& Warner 2003c) showed a strong reflection effect of
(peak-to-peak) amplitude of $\sim$ 1.1 mag, and we deduced that RXJ1039 must have been a recent nova (within
the past decade or so). As such it is the first in a class of relatively recent `overlooked' novae.
The light curves show a nearly sinusoidal intensity modulation at a period of P$_{orb}$ = 1.574 h. 

The FT (of intensities) of the entire data set, prewhitened at the orbital period and its first harmonic, 
is dominated by peaks at 1932.5 s and 721.9 s, which we associate with $\omega - \Omega$ and $2 \omega$, respectively.
There are lower amplitude peaks at the $\omega$ (P$_{spin}$ = 1444 s) and $2 \omega + \Omega$ frequencies. Only this association produces a 
sensible model (i.e., an IP with orbital sidebands and two-pole accretion) for the observed photometric modulations.
RXJ1039 is a good candidate for an extended pointed observation in the X-ray and UV region.
 \\

\begin{table}
\centering
  \caption{Candidate Intermediate Polars, optical modulations only.}
  \begin{tabular}{l r r c c c}
\\[2pt]
\hline \\ \vspace{1mm}
Name  & P$_{spin}$ (s)  & P$_{orb}$ (h) & Sidebands & Class & $<$V$>$ \\[5pt]
\hline \hline \\[1pt] 
V373 Sct  & 258.3 & & No & CN & 18.6 \\
Aqr1       & 403.7 & 3.23: & Yes & & 18.3 \\
DD Cir   & 670:  & 2.339 & No & CN & 20.1 \\
RX\,J1039.7-0507 & 1444 & 1.574 & Yes & CN: & 18.4 \\
AP Cru    & 1837 & 5.12  & No  & CN & 18.0 \\
RR Cha    & 1950 & 3.362 & No  & CN  & 18.4 \\
RX\,J0944.5+0357 & 2162 & 3.581 & No &  & 16.5 \\
V697 Sco  & 11916 & 4.49 & Yes & CN & 20.0 \\[5pt]
\hline
\end{tabular}
\label{tab1}
\newline
{\small{CN = Classical Nova, `:' denotes uncertain values.}} 
\end{table}

\section{Candidate Polars}

To date, the UCT CCD CV Survey has resulted in the detection of two probable Polars,
Nova Puppis 1991 (V351 Pup), and FIRST J102347.6+003841.\\

{\bf V351 Pup:} The light curves of Nova Puppis 1991 (V351 Pup) are so similar to V1500 Cyg (Nova Cygni 1975), that
it suggests that V351 Pup, like V1500 Cyg, is a polar. An argument in favour of this is the very
large reflection effect observed in V351 Pup (Woudt \& Warner 2001) of $\sim$ 1.2 mag 
(see also RXJ1039), that is typical of a recent nova in which the disc does not dominate 
the luminosity of the system. In the case of RXJ1039 the short orbital period implies a small 
disc; in the case of V1500 Cyg (P$_{orb}$ = 3.35 h) there is no disc. With an orbital period of
2.837 h in V351 Pup, the latter would be required, i.e., no disc, to show such large reflection effects.
G. Schmidt (private communication) has observed V351 Pup and detected no circular polarisation to
a 3$\sigma$ upper limit of 1\%, but at the time of observation V351 Pup was far less in its recovery
from its eruption than when V1500 Cyg was observed to be polarised, and hence much more
unpolarized light is probably present in V351 Pup. At V $\sim$ 19, V351 Pup will be an ideal
target for spectropolarimetric observations at large telescopes.\\

{\bf FIRST J102347.6+003841 (FIRST):} This is the first CV to be discovered from radio emission 
(Bond et al.~2002). Our photometry of FIRST of 2003 January is shown in Fig.~4. The light curves
taken by us are in sharp contrast to those shown by Bond et al.~(2002). There is some flickering
on top of what is otherwise a very regular light curve. The light curves are modulated
at a period (P$_{orb}$) of 4.75 h (the light curves shown in Fig.~4 are phased on this period). 
In the FT the fundamental and the first
harmonic of the 4.75 h periodicity are present, but no other stable period is seen (unlike RXJ1039).
The light curve has the characteristics of a reflection effect (peak-to-peak amplitude $\sim$ 0.45 mag),
so in analogy to V351 Pup and RXJ1039, FIRST could have been a recent nova (in the past few decades), that was 
overlooked. FIRST is located far from the Galactic Plane ($\ell = 243^\circ$, $b = +46^\circ$) --
only $\sim$ 10$^\circ$ away from RXJ1039 -- and could have been easily missed by nova searches. \\

\begin{figure}[t]
\plotfiddle{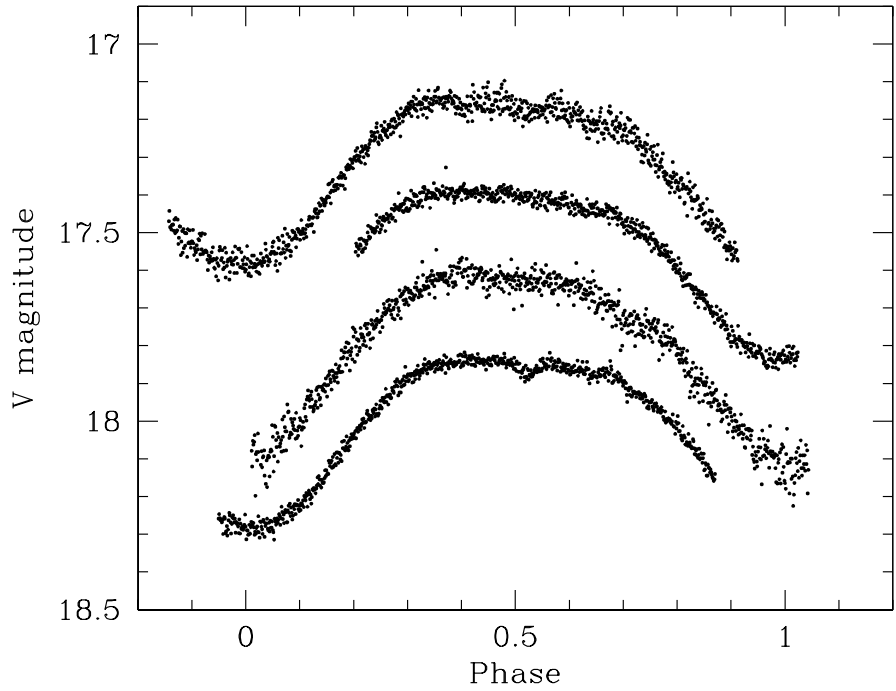}{10.5cm}{0}{65}{65}{-215}{-155}
\caption{The light curves of FIRST J102347.6+003841, phased on the orbital period of 4.75 h. 
The top light curve is at the correct brightness (V $\approx$ 17.3), the other three have been 
displaced vertically for display purposes.}
\label{woudtfig4}
\end{figure}

\bigskip
\acknowledgements{PAW is supported by strategic funds made available to BW by the University of 
Cape Town and by the National Research Foundation. BW is supported by funds from the University
of Cape Town.}

\end{document}